\documentclass[runningheads]{llncs}
\usepackage[T1]{fontenc}
\usepackage[utf8]{inputenc}

\usepackage[nounderscore]{syntax}

\usepackage{graphicx}
\usepackage{color}
\usepackage{tcolorbox}
\usepackage{listings}

\definecolor{keywordcolor}{rgb}{0.7, 0.1, 0.1}   % red
\definecolor{tacticcolor}{rgb}{0.0, 0.1, 0.6}    % blue
\definecolor{commentcolor}{rgb}{0.4, 0.4, 0.4}   % grey
\definecolor{symbolcolor}{rgb}{0.0, 0.1, 0.6}    % blue
\definecolor{sortcolor}{rgb}{0.1, 0.5, 0.1}      % green
\definecolor{attributecolor}{rgb}{0.7, 0.1, 0.1} % red

\usepackage{fontspec}
\usepackage{amsmath,amssymb,mathrsfs}
\usepackage{mathtools}

\usepackage{tikz}
\usepackage{array}
\newcolumntype{L}[1]{>{\raggedright\arraybackslash}p{#1}}

\usepackage{hyperref}

\begin{document}
\title{Verifiable Auto-Formalization of Mathematics Using a Relaxed Natural Formal Language}
\author{
Zhicheng Hui\inst{1}
\and
Lihan Xie\inst{1}
\and
Xingzhi Qi\inst{1}
\and
Zhehao Li\inst{1}
\and
Yingjun Lan\inst{1}
\and
Qinxiang Cao\inst{1}
}
\institute{Shanghai Jiao Tong University}

\maketitle
\begin{abstract}
Auto-formalization aims to translate informal mathematical content into formal languages that can be processed by theorem provers. 
However, directly targeting existing theorem provers requires LLMs to bridge a substantial representational gap between informal mathematical writing and formal proof languages.
This gap also makes semantic consistency difficult to evaluate.
We address these difficulties by introducing a Relaxed Natural Formal Language (Relaxed NFL) as an intermediate target for auto-formalization. 
The Relaxed NFL is designed to remain close to informal mathematical writing: it preserves the usual structure of informal reasoning and allows partially specified expressions and propositions, without requiring their precise interpretation to be fixed at the auto-formalization stage.
The remaining ambiguity and implicitness inherited from informal reasoning are resolved during a later elaboration stage,
which transforms Relaxed NFL proofs into Core Natural Formal Language (Core NFL) proofs with formally defined semantics.
The elaboration procedure combines rule-based transformations with LLM-generated heuristics,
while maintaining verifiability through explicit constraints on each transformation step.
The Core NFL is then used to generate proof gaps, namely verification conditions that must hold for the formalized proof to be correct.
These gaps are discharged by LLM-generated proof scripts written in a domain-specific tactic language,
which provides commands for invoking theorem libraries and domain-specific solvers implemented as part of our system.

\keywords{Auto-formalization \and Natural formal language \and Neural theorem proving.}
\end{abstract}

\section{Introduction}

Auto-formalization is the task of automatically translating informal mathematical content into a formal language that can be processed by a theorem prover, such as Rocq~\cite{10.5555/1965123}, Lean~\cite{10.1007/978-3-030-79876-5_37}, or Isabelle~\cite{10.5555/1791547}.
Because ordinary mathematical statements and proofs are written in a mixture of natural language and notation, their intended semantics may be underspecified and intermediate inferential steps are often left implicit.
Therefore, auto-formalising mathematical content into a trusted theorem prover can greatly reduce the difficulty of verifying its correctness, while also making mathematical knowledge more reusable and accessible.

Recent advances in large language models (LLMs) have led to renewed interest in auto-formalization.
Since these models are increasingly capable of processing both natural language and formal proof languages,
they can be fine-tuned or directly prompted to generate formal statements from informal ones.
Such capability can be further improved in several ways.
For instance, models can be post-trained on synthetically generated informal-formal pairs~\cite{10.5555/3737916.3740575,wang2025kiminaproverpreviewlargeformal},
augmented with retrieved formal definitions from libraries~\cite{lu2026automatedformalizationconceptualretrievalaugmented,liu2025rethinking},
or embedded in an iterative refinement loop that uses feedback from the theorem prover~\cite{guo2025autoformalizertoolfeedback,Leang_2025}.

\begin{figure}[h]
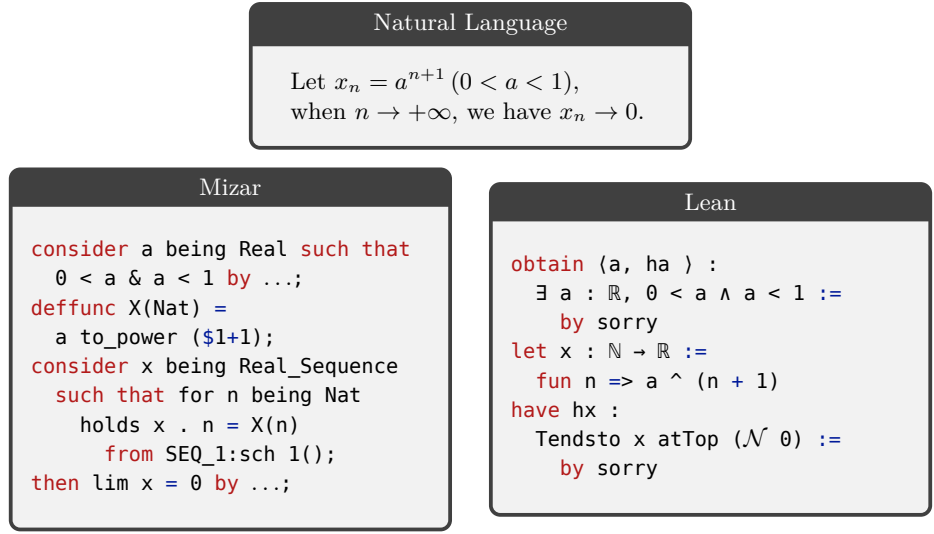

\centering

\begin{minipage}{0.48\textwidth}
\begin{tcolorbox}[
    title=Natural Language,
    fonttitle=\centering]
Let $x_n = a^{n+1} \left(0 < a < 1\right)$,

when $n \to +\infty$, we have $x_n \to 0$.
\end{tcolorbox}
\end{minipage}

\vspace{0.2cm}

\begin{minipage}{0.48\textwidth}
\begin{tcolorbox}[
    title=Mizar,
    top=2pt,
    bottom=2pt,
    left=4pt,
    right=4pt,
    fonttitle=\centering]
\begin{lstlisting}[language=lean]
consider a being Real such that
  0 < a & a < 1 by ...;
deffunc X(Nat) =
  a to_power ($1+1);
consider x being Real_Sequence
  such that for n being Nat
    holds x . n = X(n)
      from SEQ_1:sch 1();
then lim x = 0 by ...;
\end{lstlisting}
\end{tcolorbox}
\end{minipage}
\hfill
\begin{minipage}{0.48\textwidth}
\begin{tcolorbox}[
    title=Lean,
    top=2pt,
    bottom=2pt,
    left=4pt,
    right=4pt,
    fonttitle=\centering]
\begin{lstlisting}[language=lean]
obtain ⟨a, ha ⟩ :
  ∃ a : ℝ, 0 < a ∧ a < 1 :=
    by sorry
let x : ℕ → ℝ :=
  fun n => a ^ (n + 1)
have hx :
  Tendsto x atTop ((*@$\mathcal{N}$@*) 0) :=
    by sorry
\end{lstlisting}
\end{tcolorbox}
\end{minipage}

\caption{The same proof snippet written in natural language, the Mizar proof language, and the Lean proof language.}
\label{fig:discrepancy}

\end{figure}

Despite the progress, auto-formalization remains intrinsically difficult.
A major source of such difficulty is the fundamental discrepancy between informal mathematical writing and the formal languages used by existing theorem provers, as shown in Figure~\ref{fig:discrepancy}.
We can observe that the Mizar~\cite{10.1007/978-3-642-03359-9_5} and Lean proof language often require a single informal proof step to be decomposed into several explicit declarations, assumptions, and proof obligations.
For example, the informal phrase ``Let $x_n = a^{n+1} \ (0 < a < 1)$'' must be expanded into separate formal steps, including an explicit construction of the existence of such an $a$.
Similarly, the phrase ``when $n \to +\infty$'' can no longer remain a syntactic modifier attached to the statement, but must be incorporated into a formal proposition.
Besides, ordinary mathematical notation must be resolved into specific library definitions, such as the scheme \texttt{SEQ\_1:sch 1} from the Mizar Mathematical Library~\cite{10.1007/s10817-017-9440-6}, and the neighbourhood filter $\mathcal{N} \ 0$ from Mathlib~\cite{The_mathlib_Community_2020}.

This discrepancy challenges auto-formalization in two ways.
First, it requires LLMs to bridge a substantial representational gap between informal mathematical text and formal proof languages.
The scarcity of formal data further exacerbates this problem, especially for target formal languages with limited available corpora,
such as those used by proof assistants with smaller libraries or user communities, as well as domain-specific formal languages.

Second, it makes the evaluation of semantic consistency non-trivial, i.e., determining whether the generated formal statement preserves the intended meaning of the original informal statement.
Existing methods either rely on the availability of a reference formalization~\cite{liu2025generalizedtreeeditdistance,liu2025rethinking} or use LLMs or human experts as judges, making the evaluation difficult to carry out.
This issue has already had practical consequences. For example, Ospanov et al.~\cite{ospanov2026minifflean} report discrepancies between the informal and formal statements for more than half of the problems in the miniF2F benchmark~\cite{zheng2022minif2fcrosssystembenchmarkformal}, thereby compromising the reliability of some previous evaluation results.

\paragraph{Overview of the Approach} To address these difficulties at a more fundamental level, we propose using a Relaxed Natural Formal Language (Relaxed NFL) as the target language for LLM-based auto-formalization.
The Relaxed NFL is designed to remain as close as possible to natural mathematical language, so that auto-formalization into it resembles a structured normalization of informal mathematical text, rather than a full elaboration into a theorem prover language.
This reduces the reasoning burden on LLMs and makes auto-formalization largely independent of the choice of a particular formal system.

The Relaxed NFL has a formally specified syntax, which enables it to be parsed and processed by machines.
However, since it deliberately preserves part of the ambiguity, implicitness, and context dependence of informal mathematical writing,
it is not suitable to be used directly as a verification target, but only as an intermediate representation of proof.
We therefore elaborate the Relaxed NFL into a Core NFL with formally defined semantics, using a rule-based procedure supplemented by LLM-generated heuristics.
These elaboration steps isolate the discrepancy between informal mathematical writing and the target formal system, which constitutes the main difficulty of auto-formalization.
At the same time, since the input and output of each step can be formally specified, the overall process remains controllable and verifiable even when LLMs are involved.

From the Core NFL, we generate proof gaps, i.e., verification conditions whose validity is sufficient to verify the correctness of the proof.
We discharge these proof gaps by prompting LLMs to synthesize proof scripts written in a domain-specific language.
Beyond standard tactic operations such as case analysis, term rewriting, quantifier instantiation, and lemma application, the language provides commands for invoking theorem libraries and domain-specific solvers implemented as part of our system.

\begin{figure}[t]
\centering
\begin{tikzpicture}
\node (nl) at (2.2,0) {\boxed{\text{Natural Language}}};
\node (relaxed) at (2.2,-1.4) {\boxed{\text{Relaxed NFL}}};
\node (core) at (2.2,-2.8) {\boxed{\text{Core NFL}}}; 
\node (gaps) at (2.2,-4.2) {\boxed{\text{Proof Gaps}}};
\node (result) at (2.2,-5.6) {\boxed{\text{Verification Result}}};
\draw[->] (nl) -- (relaxed);
\draw[->] (relaxed) -- (core);
\draw[->] (core) -- (gaps);
\draw[->] (gaps) -- (result);
\node[align=center] at (6,-0.7) { Auto-Formalizer };
\node[align=center] at (6,-2.1) { Proof Elaborator };
\node[align=center] at (6,-3.5) { Proof Gap Generator };
\node[align=center] at (6,-4.9) { Proof Script Generator,\\ Theorem Libraries,\\ Domain-Specific Solvers,\\  };
\end{tikzpicture}
\caption{Overview of the auto-formalization pipeline.} \label{fig:pipeline}
\end{figure}
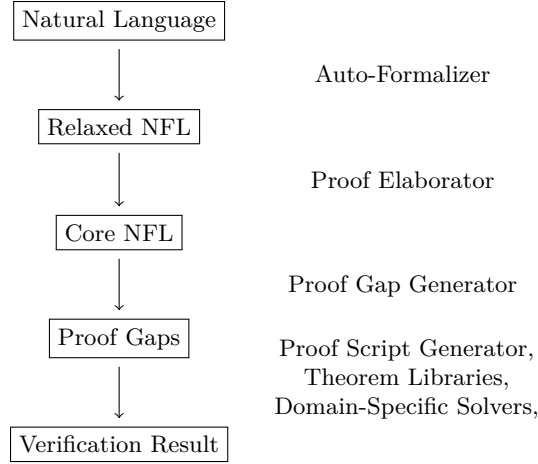

Figure~\ref{fig:pipeline} gives an overview of the entire auto-formalization pipeline.

\paragraph{Contributions}
The contributions of this paper are summarized as follows:
\begin{itemize}
\item We propose a novel approach to LLM-based auto-formalization based on an intermediate representation that resembles informal mathematical writing.
\item We study how to turn this intermediate representation into a semantically well-defined formal proof, and develop a verifiable elaboration procedure.
\item We study how proof gaps generated along this pipeline can be effectively discharged by combining LLM-guided proof search with theorem library support and domain-specific solvers.
\item We give concrete designs of our Relaxed NFL, Core NFL, and the domain-specific tactic language.
\end{itemize}

\paragraph{Organization of the Paper}
The rest of this paper is organized as follows.
Section~2 presents the Relaxed and the Core NFL, as well as the LLM-based auto-formalization process into Relaxed NFL.
Section~3 demonstrates the elaboration of the Relaxed NFL into the Core NFL via three representative steps: notation resolution, problem-solving translation, and implicit variable scope resolution.
Section~4 explains how proof gaps are generated from the Core NFL and subsequently discharged.
Section~5 evaluates whether LLMs can reliably generate Relaxed NFL proofs under different settings.
Finally, Sections~6 and~7 are devoted to related work and conclusions, respectively.

\section{Natural Formal Language}

In principle, natural formal language is intended to be extensible and can be instantiated in different mathematical domains.
The particular natural formal languages presented in this paper are designed with reference to the proof structures, notation, and mathematical knowledge appearing in the textbook \textit{Problems in Mathematical Analysis} by B.~P.~Demidovich~\cite{demidovich1970problems}.

The remainder of this section is organized as follows.
We first briefly introduce the Relaxed NFL, then describe how informal mathematical proofs can be auto-formalized into the Relaxed NFL by using LLMs.
Finally, we present the Core NFL, whose formal semantics is defined by a proof-gap generation function.

\subsection{The Relaxed Natural Formal Language}

The Relaxed NFL remains close to informal mathematical writing in two respects.
First, it preserves the common reasoning patterns of informal proofs, so that the overall proof structure is largely maintained after auto-formalization.
Second, it allows partially specified terms and propositions by retaining informal mathematical writing without requiring every expression to be resolved to a precise definition in a specific library.

As an example, Appendix~\ref{app:A} presents the Relaxed NFL formalization of an IMO functional-equation problem.
As shown in this example, the Relaxed NFL is a declarative-style proof language.
Among other things, user can introduce new symbols, derive intermediate conclusions, reason locally under assumptions, and solve equations.

The Relaxed NFL serves only as a surface syntax for LLM-based auto-formalization.
Terms in the Relaxed NFL do not yet have fully precise mathematical meanings at this stage.
Some notations may remain ambiguous, and information on variable scopes may be left implicit, as shown by the occurrence of free variables in expressions.
These partially specified expressions are later resolved into precise mathematical objects through elaboration, as described in Section~3.

The detailed grammar of the Relaxed NFL is given in Appendix~\ref{app:B}.
The non-terminal \texttt{<program>} can be either a proof problem or a solving problem.
They share the same reasoning steps, but are translated into the Core NFL differently.
The non-terminal \texttt{<term>} represents mathematical propositions and expressions.
We omit its grammar here
and introduce the relevant fragments in Section~3 when needed.

\subsection{Auto-Formalization into the Relaxed NFL}

We instruct LLMs to translate informal mathematical content into the Relaxed NFL in two ways: in-context learning and fine-tuning.

For in-context learning, the prompt first provides the complete grammar of the Relaxed NFL, followed by 16 examples that serve as few-shot demonstrations.
The model is then asked to translate the input into the target language.
We then use the obtained informal-formal pairs as supervised training data, and perform LoRA fine-tuning~\cite{hu2021loralowrankadaptationlarge} so that the model learns the grammar of the Relaxed NFL more directly.

\subsection{The Core Natural Formal Language}

The Core NFL is an internal proof language obtained by elaborating the surface-level constructs of the Relaxed NFL into a small number of primitive proof steps over formally specified terms.
Unlike the Relaxed NFL, the Core NFL has a precisely defined semantics, and captures the information necessary for generating proof gaps.

We define the abstract syntax of the Core NFL as follows.
We assume two syntactic categories:
\(\mathsf{Term}\), whose elements represent mathematical expressions and propositions in the Core NFL,
and \(\mathsf{Method}\), whose elements represent proof methods,
with a distinguished element \(\varnothing_{\mathsf{M}}\) denoting the empty method.
We write \(\mathsf{Term}^{*}\) for the set of finite, possibly empty sequences of terms.

Based on this, we define the set of proof states \(\mathsf{PState}\) by
\[
  \mathsf{PState} = \mathsf{Term}^{*} \times \mathsf{Term}.
\]
We write a proof state consisting of a sequence of assumptions \(\Gamma \in \mathsf{Term}^{*}\) and a proof goal \(G \in \mathsf{Term}\) in sequent form as \(\Gamma \vdash G\).

Let \(\mathsf{Proof}\) be the least set satisfying
\[
\begin{aligned}
\mathsf{Proof}
= \ & \{\mathsf{Forward}(M,P,\pi)
      \mid M \in \mathsf{Method},\ P \in \mathsf{Term},\ \pi \in \mathsf{Proof}\} \\
\cup \ & \{\mathsf{Backward}(M,P,\pi)
      \mid M \in \mathsf{Method},\ P \in \mathsf{Term},\ \pi \in \mathsf{Proof}\} \\
\cup \ & \{\mathsf{Subgoal}(P,\pi_1,\pi_2)
      \mid P \in \mathsf{Term},\ \pi_1,\pi_2 \in \mathsf{Proof}\} \\
\cup \ & \{\mathsf{QED}\}.
\end{aligned}
\]
The set of Core NFL program \(\mathsf{Program}\) is then defined by
\[
\mathsf{Program}
=
\mathsf{PState} \times \mathsf{Proof}.
\]

The constructors of \(\mathsf{Proof}\) are interpreted as follows:
\begin{itemize}
  \item \(\mathsf{Forward}(M,P,\pi)\) derives the proposition \(P\) from the current proof state using the method \(M\), adds it to the available assumptions, and then continues with the proof \(\pi\).
  \item \(\mathsf{Backward}(M,P,\pi)\) reduces the current proof goal to the proposition \(P\) using the method \(M\), and then continues with the proof \(\pi\).
  \item \(\mathsf{Subgoal}(P,\pi_1,\pi_2)\) introduces the proposition \(P\) as an intermediate subgoal, proves it by \(\pi_1\), adds it to the available assumptions, and then continues with the proof \(\pi_2\).
  \item \(\mathsf{QED}\) terminates the proof, indicating that the current proof goal is derivable from the available assumptions.
\end{itemize}

A Core NFL program is therefore a pair \(\mathcal{P} = (\sigma,\pi)\), where
\(\sigma \in \mathsf{PState}\) represents the current proof state,
and \(\pi \in \mathsf{Proof}\) represents the proof to be executed from that state.

\section{Elaborating the Relaxed NFL into the Core NFL}

In this section, we present how the Relaxed NFL is elaborated into the Core NFL in a verifiable way.
Such elaboration generally cannot be achieved by purely syntactic transformation of the proof.
Instead, we often have to perform semantic analysis in order to recover information that is left implicit.
Some of the simple cases can be formulated as deterministic rules,
while others require a closer analysis of the proof and may even involve inferring the author's intended meaning.
In such cases, we use LLMs to give heuristics, while keeping the elaboration step verifiable by placing models inside a suitable harness.
For example, the model may be restricted to choosing among a finite set of candidates generated by deterministic rules, or its output may be required to pass a dedicated checker before being accepted.

Since different mathematical domains use different notations and proof writing conventions, and tolerate different degrees of implicitness, the concrete elaboration process is dependent on the domain.
In this section, we focus on mathematical analysis and present three representative elaboration steps extracted from our implementation.
They exemplify three different levels at which ambiguity and implicitness arise in informal mathematical writing: the term level, the proof-step level, and the proof-state level.

Because the Core NFL can be regarded as a restricted fragment of the Relaxed NFL,
we present the elaboration steps in this section as transformations on the Relaxed NFL itself.
Each step resolves certain ambiguity or implicitness,
thereby moving the proof closer to the Core NFL fragment.

\subsection{Notation Resolution}

A common source of ambiguity in informal mathematical writing is notation overloading,
where the same notation may denote different mathematical objects.
A simple example is the notation \(|x|\).
Depending on the meaning of \(x\), \(|x|\) may denote the absolute value of a real number, or the cardinality of a set.
Consequently, the surface syntax \(|x|\) in the Relaxed NFL does not automatically resolve to a Core NFL term by itself.

In order to perform notation resolution, we first need to augment the definition of terms in the Relaxed NFL with the following constructs:

\vspace{0.4cm}
\begin{list}{}{%
  \setlength{\leftmargin}{0.33\linewidth}%
  \setlength{\rightmargin}{0.33\linewidth}%
  \setlength{\topsep}{0pt}%
  \setlength{\partopsep}{0pt}%
  \setlength{\parsep}{0pt}%
  \setlength{\itemsep}{0pt}%
}
\item\relax
\begin{grammar}
<term> ::= ...
\alt "|" <term> "|"
\alt "IsReal(" <term> ")"
\alt "IsSet(" <term> ")"
\alt "Abs(" <term> ")"
\alt "Card(" <term> ")"
\end{grammar}
\end{list}
\vspace{0.4cm}

Here, \(|\cdot|\) is the overloaded surface notation, while \(\mathsf{Abs}(\cdot)\) and \(\mathsf{Card}(\cdot)\) are disambiguated forms that can be deterministically translated into the corresponding Core NFL terms, namely the absolute value of a real number and the cardinality of a set.
The predicates \(\mathsf{IsReal}\) and \(\mathsf{IsSet}\) record information about whether a term has been treated as a real number or as a set.
They are not part of a static type system for Relaxed NFL, but annotations used to guide the interpretation of overloaded notation.
In this respect, they are similar in spirit to Mizar's soft type system~\cite{10.5555/1792233.1792261}.

We perform a static analysis over the Relaxed NFL and maintain an environment that stores such annotations. These annotations may come from two sources.

First, they may be introduced by the LLM during the initial auto-formalization process.
The prompts are designed to encourage the LLM to preserve explicit information from the informal proof about what kind of mathematical object a term denotes.
For example, the informal phrase ``for every real number \(t\)'' is supposed to be translated into the Relaxed NFL by introducing \(t\) together with the annotation \(\mathsf{IsReal}(t)\).

Second, our system may infer them by constraint propagation over such annotations.
It derives constraints from operator signatures and records the corresponding annotations in the environment.
For example, if an operator expects a set argument and is applied to \(t\), the constraint \(\mathsf{IsSet}(t)\) is generated.

After constraint propagation, the elaborator checks each unresolved overloaded notation.
When the environment determines a unique interpretation, the notation is replaced by the corresponding disambiguated form.
For instance, \(|t|\) is resolved as \(\mathsf{Abs}(t)\) if \(\mathsf{IsReal}(t)\) is available, and as \(\mathsf{Card}(t)\) if \(\mathsf{IsSet}(t)\) is available.
Otherwise, the system queries the LLM with instructions, the surrounding proof context, and the finite set of remaining candidate interpretations, asking it to select the intended one.

\subsection{Problem-Solving Translation}

In addition to theorem proving, the Relaxed NFL also supports problem-solving.
Problem-solving does not merely require proving a given proposition.
Rather, it requires constructing an answer together with a proof that the answer is correct~\cite{liu2025theoremprovingformulationframework,10.1145/3759425.3763384}.
For example, consider a Relaxed NFL program of the form:
\[
\texttt{Find all x such that P(x) \{ ... \}}
\]

When it terminates, we should be able to identify a candidate solution set \(S\) and prove that \(S\) is exactly the set of all solutions.
Once \(S\) has been determined, the ending of the problem-solving step is reduced to the following step:
\[
\texttt{We have forall (x), x in S <==> P(x)}
\]

However, the candidate solution set \(S\) cannot be recovered at the syntactic level alone.
This is because the reasoning steps of problem-solving may not all preserve logical equivalence with the original constraint \(P(x)\).
Some may only derive a sufficient or necessary condition,
and such steps are not distinguishable from their surface syntax alone.
Therefore, identifying the exact solution set is a semantic task, as it depends on whether the derived candidates genuinely satisfy the original constraint.

This task is well suited to LLMs, because the model only needs to propose a candidate answer, whose validity can be verified symbolically.
Thus, we use in-context learning to infer the exact candidate solution set for each problem-solving task by providing the model with instructions and the surrounding proof context.
Other kinds of problem-solving steps are handled analogously.

In some cases, we explicitly check whether the candidates obtained satisfy the constraints.
For example, consider a Relaxed NFL program of the form:
\[
\begin{array}{l}
\texttt{Solve the system of equations P(x), Q(x) \{} \\
\texttt{\ \ ... After checking, x = x0 satisfies the constraints} \\
\texttt{\}}
\end{array}
\]
This can be translated into a step asserting that the checked candidate satisfies the original equations:
\[
\begin{array}{l}
\texttt{Solve the system of equations P(x), Q(x) \{} \\
\texttt{\ \ ... We have P(x0) and Q(x0)} \\
\texttt{\}}
\end{array}
\]

\subsection{Implicit Variable Scope Resolution}

Informal mathematical writing often omits parts of the information on the scope of variables.
In particular, adjacent steps may share locally introduced variables and conditions without explicitly specifying their scope.
For example, consider the following simplified Relaxed NFL proof adapted from Appendix~\ref{app:A}:

\begin{tcolorbox}
\begin{lstlisting}[keepspaces=true]
Solve: Find all f such that
  ...,
  forall (x, y in Real),
    f(x - f(y)) = f(f(y)) + x * f(y) + f(x) - 1

Solution:
   Set f(0) = c
   ...
1. We have f(x) = frac(c + 1, 2) - frac(x ^ 2, 2)
2. [@method By x ^ 2 = (-x) ^ 2 @]
   We have f(x) = f(-x)
3. We have f(c) = f(-c)
4. We have f(c) = f(c) + c - 1
5. We have c = 1
6. We have f(x) = 1 - frac(x ^ 2, 2)
\end{lstlisting}
\end{tcolorbox}

Here, the same name \(x\) occurs free in Steps 1, 2, and 6,
suggesting that these occurrences may refer to the same variable
and requiring the current proof context to be interpreted within the scope of \(x\),
which has two semantic implications.

First, variable scopes determine the interpretation of a term written in the Relaxed NFL.
The declaration \texttt{forall (x, y in Real)} at the beginning of the problem provides evidence that its proof steps may enter the scope of a universally quantified real variable \(x\).
Thus, Step 1 should be interpreted as deriving the closed proposition \(\forall x \in \mathbb{R},\, f(x)=\frac{c+1}{2}-\frac{x^2}{2}\).
Similarly, Step 2 should be interpreted as deriving \(\forall x \in \mathbb{R},\, f(x)=f(-x)\).
In this sense, the interpretation of a surface syntax term is determined by the current variable scope.

One may be tempted to think that this interpretation can always be obtained simply by closing the free variables syntactically occurring in the term.
We show in the following paragraphs that this is not the case.

Second, variable scopes may also suggest how a proposition is derived,
which in turn affects the interpretation of a term written in the Relaxed NFL.
For example, for an arbitrarily fixed \(x \in \mathbb{R}\), Step 1 gives \(f(x)=\frac{c+1}{2}-\frac{x^2}{2}\).
Together with Step 5, namely \(c=1\), it yields Step 6, namely \(f(x)=1-\frac{x^2}{2}\).
Since the interpretation of Step 6 is still deriving the universal proposition \(\forall x \in \mathbb{R},\, f(x)=1-\frac{x^2}{2}\),
rather than properties about one particular \(x\) in the current scope.
What happens here is that, when \(c=1\), we prove
\(\forall x \in \mathbb{R},\, f(x)=\frac{c+1}{2}-\frac{x^2}{2}\)
\(\Rightarrow\)
\(f(x)=1-\frac{x^2}{2}\)
in order to justify
\((\forall x \in \mathbb{R},\, f(x)=\frac{c+1}{2}-\frac{x^2}{2})\)
\(\Rightarrow\)
\((\forall x \in \mathbb{R},\, f(x)=1-\frac{x^2}{2})\),
which constitutes a form of local derivation carried out within the scope of \(x\).

However, being in the same scope for \(x\) does not mean that the derivation is always local in the above sense.
For example, in deriving Step 2 from Step 1, what is proved is the implication
\((\forall x \in \mathbb{R},\, f(x)=\frac{c+1}{2}-\frac{x^2}{2})\)
\(\Rightarrow\)
\((\forall x \in \mathbb{R},\, f(x)=f(-x))\)
itself, rather than
\(\forall x \in \mathbb{R},\, f(x)=\frac{c+1}{2}-\frac{x^2}{2}\)
\(\Rightarrow\)
\(f(x)=f(-x)\),
which in fact does not hold.

Similarly, when multiple variables are present in the scope,
some of them may not occur freely in the derived surface syntax term,
but a non-local derivation as shown in the previous paragraph will still make the interpretation to be quantified by these variables.
It demonstrates that determining the precise interpretation of surface syntax terms which contains free variables involves a semantic analysis as to the actual reasoning pattern of the proof (local/non-local), instead of a syntactic analysis of free variables.
Consequently, we delegate the resolution of implicit variable scopes to LLMs.

In order to perform implicit variable scope resolution, we first need to augment definition of proofs and terms in the Relaxed NFL with the following constructs:

\vspace{0.4cm}
\begin{list}{}{%
  \setlength{\leftmargin}{0.08\linewidth}%
  \setlength{\rightmargin}{0.08\linewidth}%
  \setlength{\topsep}{0pt}%
  \setlength{\partopsep}{0pt}%
  \setlength{\parsep}{0pt}%
  \setlength{\itemsep}{0pt}%
}
\item\relax
\begin{grammar}
<local\_action\_step> ::= ...
\alt "[@scope" ("["$\forall$ <term> "]"|"["$\exists$ <term> "]"|"[" <term> "]")* "@]"

<term> ::= ...
\alt "[@free" <term> ("," <term>)* "@]"
\end{grammar}
\end{list}
\vspace{0.4cm}

The construct \texttt{[@scope ... @]} allows the LLM to explicitly annotate the implicit variable scope via \texttt{<local\_action\_step>} in the Relaxed NFL. 
The construct \texttt{[@free ... @]} marks the free variables occurring in a term, it is precomputed and annotated to the Relaxed NFL proof given to the LLM, helping it to infer the implicit scope covering the step.

The output of this elaboration step is then verified by a deterministic checker.
The checker ensures that all the annotated free variables occur within a matching scope,
and that no other part of the proof has been modified.

\begin{figure}[t]
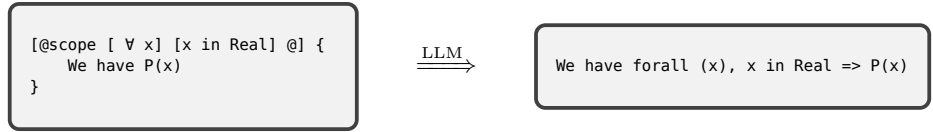

\centering
\begin{minipage}{0.38\textwidth}

\begin{tcolorbox}[
  top=2pt,
  bottom=2pt,
  left=4pt,
  right=4pt
]
\begin{lstlisting}[
  basicstyle=\ttfamily\scriptsize,
  columns=fullflexible,
  keepspaces=true
]
[@scope [ ∀ x] [x in Real] @] {
    We have P(x)
}
\end{lstlisting}
\end{tcolorbox}
\end{minipage}
\hfill
\begin{minipage}{0.08\textwidth}
\centering
\(\xRightarrow{\text{LLM}}\)
\end{minipage}
\hfill
\begin{minipage}{0.43\textwidth}
\begin{tcolorbox}[
  top=2pt,
  bottom=2pt,
  left=4pt,
  right=4pt
]
\begin{lstlisting}[
  basicstyle=\ttfamily\scriptsize,
  columns=fullflexible,
  keepspaces=true
]
We have forall (x), x in Real => P(x)
\end{lstlisting}
\end{tcolorbox}
\end{minipage}

\caption{Elimination of the variable scope annotation in LLM's output.}
\label{fig:scope}
\end{figure}

Finally, we eliminate the explicit variable scope annotation in the Relaxed NFL as in Figure~\ref{fig:scope}.
This is done by adding the content of variable scope annotation to every terms in the scope as quantifiers and premises.

\section{Generating and Discharging Proof Gaps}

In this section, we define the semantics of Core NFL programs.
Rather than interpreting an element of \(\mathsf{Program}\) as a complete formal proof,
we interpret it as a procedure that generates a collection of proof gaps.
These proof gaps must subsequently be discharged by a trusted backend in order for the whole proof to be certified.

\subsection{The Proof Gap Generator}

A proof gap consists of a proof method together with a proof state.
Formally, we define the set of proof gaps \(\mathsf{Gap}\) by:
\[
  \mathsf{Gap} = \mathsf{PState} \times \mathsf{Method}.
\]
We write a proof gap consisting of a proof
state \((\Gamma,G) \in \mathsf{PState}\) and method \(M \in \mathsf{Method}\) in sequent form as \(\Gamma \vdash_{M} G\).
It represents the obligation of deriving the
goal \(G\) from the available assumptions \(\Gamma\) using the method \(M\).

We denote by \(\Gamma\,P\) the sequence obtained by appending \(P\) to the end of \(\Gamma\).
The proof gap generator for the Core NFL is a function
\[
  \mathsf{PGG}
  :
  \mathsf{PState} \times \mathsf{Proof}
  \longrightarrow
  \mathcal{P}_{\mathrm{fin}}(\mathsf{Gap}),
\]
defined by structural recursion on the proof as follows:
\[
\begin{alignedat}{2}
&\mathsf{PGG}(\Gamma \vdash G,\mathsf{Forward}(M,P,\pi))
&\ = {}& \{\Gamma \vdash_{M} P\}
\cup \mathsf{PGG}(\Gamma\,P \vdash G,\pi),
\\[0.6em]
&\mathsf{PGG}(\Gamma \vdash G,\mathsf{Backward}(M,P,\pi))
&\ = {}& \{\Gamma\,P \vdash_{M} G\}
\cup \mathsf{PGG}(\Gamma \vdash P,\pi),
\\[0.6em]
&\mathsf{PGG}(\Gamma \vdash G,\mathsf{Subgoal}(P,\pi_1,\pi_2))
&\ = {}& \mathsf{PGG}(\Gamma \vdash P,\pi_1)
\cup \mathsf{PGG}(\Gamma\,P \vdash G,\pi_2),
\\[0.6em]
&\mathsf{PGG}(\Gamma \vdash G,\mathsf{QED})
&\ = {}& \{\Gamma \vdash_{\varnothing_{\mathsf{M}}} G\}.
\end{alignedat}
\]

\subsection{Discharging Generated Proof Gaps}

A Core NFL program is certified by checking all generated proof gaps. We define two functions
\[
  \mathsf{CheckGap} : \mathsf{Gap} \to \{\bot,\top\},
  \qquad
  \mathsf{CheckProgram} : \mathsf{Program} \to \{\bot,\top\},
\]
such that
\[
  \forall \, (\sigma,\pi) \in \mathsf{Program},\,
  \mathsf{CheckProgram}(\sigma,\pi)
  =
  \bigwedge_{\mathcal{G} \in \mathsf{PGG}(\sigma,\pi)}
  \mathsf{CheckGap}(\mathcal{G}).
\]
Thus, for a Core NFL program \(\mathcal{P}\), \(\mathsf{CheckProgram}(\mathcal{P}) = \top\) means that all the generated proof gaps are successfully discharged.

% 由于我们的Core NFL继承了自然语言的推理结构，因此，proof gap可能会比较大，只靠自动定理证明方法不足够。
% 为了解决这个问题，我们在这一步利用LLM来指导proof gap的discharge。
% 具体地来说，我们设计了一个作用在proof gap上的tactic language，它可以让LLM写proof script完成proof gap的证明。
% tactic language展示如下

\subsubsection{Tactic Language for Proof Gaps}

Since the Core NFL inherits the reasoning structure of the informal proof,
the generated proof gaps may still contain non-trivial intermediate reasoning.
As a result, relying only on automatic theorem proving techniques is often insufficient.

To address this issue, we use LLMs to facilitate the proof search.
Concretely, we design a domain-specific tactic language that operates on individual proof gaps.
Given a proof gap, the LLM is asked to synthesize a proof script in this language,
which specifies a sequence of commands that either close the gap or reduce it to simpler subgaps.
The method annotations in the Relaxed NFL, corresponding to the non-terminal \texttt{<method>} in Appendix~\ref{app:B}, are used to provide proof search hints.

The main commands of the tactic language are summarized in Table~\ref{tab:tactic}.
Most of them correspond to standard proof-state transformations in tactic-based proof languages.
The last three commands are more special:
\texttt{auto\_solve} and \texttt{auto\_replace} provides a higher level of automation by invoking our domain-specific solvers to close proof gaps directly or replace a term with a semantically equivalent one.
\texttt{get\_prop} retrieves and instantiates propositions from our theorem libraries,
allowing the model to use established theorems without reconstructing them from scratch.

\begin{table}[t]
\centering
\caption{Main commands of the tactic language used for discharging proof gaps.}
\label{tab:tactic}
\begin{tabular}{L{2.4cm}L{9.6cm}}
\hline
Command & Use \\
\hline
\texttt{assert} & Proves a proposition and adds it to the assumptions. \\
\texttt{apply} & Reduces the proof goal using a matching implicational assumption. \\
\texttt{case\_analysis} & Splits a proof gap according to a disjunctive assumption. \\
\texttt{rewrite} & Rewrites assumptions or the proof goal by equalities or equivalences. \\
\texttt{get\_forall} & Instantiates universal quantifiers in the proof goal with fresh variables. \\
\texttt{get\_condition} & Introduces premises of an implicational proof goal as assumptions. \\
\texttt{get\_exists} & Eliminates existential quantifiers in an assumption using fresh variables. \\
\texttt{exists} & Instantiates existential quantifiers in the proof goal with witnesses. \\
\texttt{use\_condition} & Specializes an implicational assumption using matching premises. \\
\texttt{auto\_replace} & Performs semantic term transformation. \\
\texttt{auto\_solve} & Invokes automatic solvers on selected assumptions. \\
\texttt{get\_prop} & Instantiates hypotheses or imports theorem-library facts. \\

\texttt{auto\_solve} & Invokes solvers to automatically close the proof gap. \\
\texttt{auto\_replace} & Invokes solvers to perform semantically-equivalent term replacement. \\
\texttt{get\_prop} & Retrieves and instantiates a proposition from the theorem library. \\

\hline
\end{tabular}
\end{table}

\subsubsection{Theorem Libraries and Domain-Specific Solvers}

The theorem libraries contain formalized propositions extracted from \textit{Problems in Mathematical Analysis}, together with commonly used facts about sets, inequalities, elementary functions, limits, etc.

The domain-specific solvers are designed to discharge recurring proof gap patterns for which effective decision procedures are available.
Rather than being used in isolation, these solvers are composed automatically.
Each solver may close a proof gap directly or generate new subgaps, which can in turn be handled by other solvers.
This is implemented as a breadth-first search.
The main solvers are summarized in Table~\ref{tab:solvers}.

\begin{table}[t]
\centering
\caption{Main domain-specific solvers used for discharging proof gaps.}
\label{tab:solvers}
\begin{tabular}{L{3.0cm}L{9.0cm}}
\hline
Solver & Use \\
\hline
\texttt{EqbSolver}
& Reflexivity and exact matching against assumptions. \\
\texttt{PolyRatSolver}
& Normalization of polynomial and rational expressions. \\
\texttt{RootFracSolver}
& Normalization of radical and fractional expressions. \\
\texttt{SgnSolver}
& Inequality reasoning using basic facts about sign. \\
\texttt{IntervalSolver}
& Propagation of interval constraints. \\
\texttt{SmtLraSolver}
& Linear real arithmetic theory reasoning. \\
\texttt{DerivSolver}
& Calculation of symbolic derivatives.  \\
\texttt{IntegralSolver}
& Calculation of symbolic integrations. \\
\texttt{ConDiffFuncSolver}
& Continuity and differentiability of functions. \\
\texttt{ConstTypeSolver}
& Membership of constant expressions in standard number sets. \\
\texttt{NotUnfoldSolver}
& Normalization of negated propositions. \\
\hline
\end{tabular}
\end{table}

\section{Evaluation}

We evaluate the learnability of the Relaxed NFL as a target language for LLM-based auto-formalization.
We report pass@1 and pass@3.
An auto-formalization attempt is counted as a pass if it can semantically yield semantically well-defined proof gaps,
while the discharge of these proof gaps is not included in this evaluation.
For each problem, we sample three independent candidates from the model.
The pass@1 rate is the proportion of problems for which the first generated candidate is counted as a pass,
while the pass@3 rate is the proportion of problems for which at least one of the three generated candidates is counted as a pass.

Our dataset contains 3600 problems extracted by OCR from \textit{Problems in Mathematical Analysis},
covering major topics in elementary mathematical analysis,
including limits, continuity, differentiation, integration, series, and multivariable calculus.

We first use Qwen3-235B-A22B-Instruct~\cite{yang2025qwen3technicalreport}, a large MoE model with 235B total parameters and 22B activated parameters, for the few-shot auto-formalization described in Section~2.2.
Without chain-of-thought prompting (CoT), the model achieves 44.4\% pass@1 and 72.3\% pass@3.
With CoT, the rates increase to 70.3\% pass@1 and 90.4\% pass@3.

We then select 2,000 correctly generated informal-formal pairs from this stage as supervised training data, and use the remaining examples as the test set.
Finally, we fine-tune Qwen2.5-7B-Instruct~\cite{yang2024qwen2technicalreport},
a smaller dense model from an earlier generation of the Qwen series, with LoRA, using rank \(r=128\), scaling factor \(\alpha=256\), and learning rate \(\eta=10^{-4}\).
The fine-tuned model achieves 83.6\% pass@1 and 89.9\% pass@3.
The results are summarized in Table~\ref{tab:evaluation}.

\begin{table}[t]
\centering
\caption{Evaluation of Relaxed NFL generation.}
\label{tab:evaluation}
\begin{tabular}{L{4.2cm}L{5.1cm}L{1.2cm}L{1.2cm}}
\hline
Model & Setting & pass@1 & pass@3 \\
\hline
Qwen3-235B-A22B-Instruct
& Few-shot prompting without CoT
& 44.4\% & 72.3\% \\
Qwen3-235B-A22B-Instruct
& Few-shot prompting with CoT
& 70.3\% & 90.4\% \\
Qwen2.5-7B-Instruct
& LoRA fine-tuning on 2,000 informal-formal pairs
& 83.6\% & 89.9\% \\
\hline
\end{tabular}
\end{table}

These results show that Relaxed NFL proofs can be reliably generated by LLMs.
In the few-shot prompting experiments, we observe that many failed samples are caused by the model drifting toward other formal languages, such as LaTeX-like notation.
This suggests that these failures mainly reflect the model's prior bias toward familiar mathematical syntaxes,
rather than an intrinsic difficulty of learning the Relaxed NFL grammar.
When the model is allowed to reason explicitly about the target grammar, it is less likely to drift away from Relaxed NFL,
as suggested by the improvement brought by CoT.

More importantly, the fine-tuning result shows that the Relaxed NFL can be learned by smaller models with a lightweight training method and limited training data.
Using only 2,000 informal-formal pairs and LoRA fine-tuning,
Qwen2.5-7B-Instruct achieves a higher pass@1 rate than the much larger Qwen3-235B-A22B-Instruct with CoT, and a comparable pass@3 rate.
This suggests that a suitable target language can substantially reduce the difficulty of LLM-based auto-formalization.

\section{Related Work}

\paragraph{Theorem-Proving Large Language Models}
One important role of auto-formalization is to synthesize large amounts of formal proof data.
Since the available formal proof corpora are relatively scarce, auto-formalization has become an essential component in the training pipelines of theorem proving LLMs, i.e., LLMs whose primary goal is to prove theorems in proof assistants.
For example, the post-training data of DeepSeek-Prover~\cite{xin2024deepseekproveradvancingtheoremproving}, Goedel-Prover~\cite{lin2025goedelproverfrontiermodelopensource}, and AlphaProof~\cite{Hubert2025} all contain a large number of auto-formalized Lean 4 proofs.

However, existing work on theorem-proving language models mainly focuses on exploring more effective post-training methods and proof-search strategies.
For instance, DeepSeek-Prover-V2~\cite{ren2025deepseekproverv2advancingformalmathematical} uses reinforcement learning to help the model decompose a problem into subgoals.
Other systems improve the quality of the feedback available to the model.
For instance, Seed-Prover~\cite{chen2025seedproverdeepbroadreasoning} combines Lean compiler feedback, proved lemmas, and self-summarization to refine generated proofs,
while LEGO-Prover~\cite{wang2023legoproverneuraltheoremproving} maintains a growing library of verified lemmas and uses them as skills during inference.
In contrast, relatively less attention has been paid to whether the reasoning ability of LLMs can be improved by carefully designing the target formal representation itself, 
or to what kind of language is suitable for mathematical reasoning by LLMs while still remaining machine-checkable.
In this respect, our work is complementary to these systems.

\paragraph{Natural-Language-Like Formal Languages}

Proof languages that aim to resemble natural language, or that take readability as a design goal, are usually described as declarative-style proof languages.
Unlike term-style or procedural-style proof languages, they allow users to write proofs by explicitly stating assumptions, intermediate claims, and local arguments, which are common structures in informal mathematical writing. Many theorem provers support such declarative style.
For example, Isabelle's Isar~\cite{wenzel1999isar} language represents proofs as structured mathematical texts.
It supports nested proof contexts and the introduction of local variables, but its proof contexts are still tied to the current proof obligation,
and local variables and assumptions cannot be introduced with the same flexibility as in informal mathematical writing.

Lean's verbose-lean4 language~\cite{massot:LIPIcs.ITP.2024.27} combines declarative and procedural styles.
It provides declarative constructs such as \texttt{have}, while also allowing users to manipulate the proof state through a more natural-language-like surface syntax that is subsequently elaborated into tactics.
However, it has limited support for structured proofs.
For example, nested subproofs are mainly expressed through \texttt{by}-blocks, whose contents must be written as tactics rather than in the same declarative style.
The Mizar proof language~\cite{10.1007/978-3-642-03359-9_5} and Naproche's ForTheL~\cite{10.1007/978-3-030-79876-5_36} also follow a declarative style close to conventional mathematical prose.
Previous work has shown that such natural-language-like proof languages can improve LLM-based auto-formalization.
For instance, Draft, Sketch, and Prove~\cite{jiang2023draftsketchproveguiding} uses natural-language drafts to generate Isar proof skeletons.

The Relaxed NFL presented in this paper not only supports structured proofs organized by introducing local variables and assumptions, but also has richer structures such as problem-solving.
In this respect, the Relaxed NFL is closer to informal mathematical writing.
More importantly, terms and propositions in existing proof languages usually have to refer to precise library definitions, and do not allow partially specified expressions or the occurrence of free variables,
while the Relaxed NFL preserve the surface form of informal mathematical notation, allowing a degree of ambiguity and context dependence.
Subsequent elaboration steps can repair term-level ambiguities and explicitly reconstruct implicit local proof contexts to eliminate free variables,
thereby offloading part of the reasoning burden from LLMs during auto-formalization.

\section{Conclusion}

In this paper, we present an auto-formalization pipeline based on a relaxed natural formal language. The main idea is to use an intermediate representation that remains close to informal mathematical writing, instead of directly instructing LLMs to generate the formal proof language of a specific theorem prover. This design reduces the representational gap that LLMs need to bridge and makes auto-formalization less dependent on the details of a particular formal system.

We introduce the Relaxed NFL as a surface language for LLM-based auto-formalization, and the Core NFL as a semantically well-defined proof language.
We then describe a verifiable elaboration pipeline from the Relaxed NFL to the Core NFL, taking notation resolution, problem-solving translation, and implicit variable scope resolution as examples.

To verify the correctness of proofs, we generate proof gaps from the Core NFL and design a dedicated domain-specific tactic language in which LLMs can generate proof scripts to discharge these gaps.
We incorporate theorem libraries and domain-specific solvers as part of our system to enhance the tactic language and facilitate LLM-based proof gap discharging.

Our work demonstrates the feasibility and benefits of using a proof language closer to natural mathematical writing for auto-formalization.
With a carefully designed elaboration process, the logical rigor that is intentionally left implicit in the relaxed proof language can be recovered in a controlled and verifiable way, thereby supporting proof verification.
Such design not only makes the language easier for LLMs to acquire and use, but also improves the readability of formal proofs.

\bibliographystyle{splncs04}
\bibliography{reference}

@book{10.5555/1965123,
author = {Bertot, Yves and Castran, Pierre},
title = {Interactive Theorem Proving and Program Development: Coq'Art The Calculus of Inductive Constructions},
year = {2010},
isbn = {3642058809},
publisher = {Springer Publishing Company, Incorporated},
edition = {1st}
}

@book{10.5555/1791547,
author = {Nipkow, Tobias and Wenzel, Markus and Paulson, Lawrence C.},
title = {Isabelle/HOL: a proof assistant for higher-order logic},
year = {2002},
isbn = {3540433767},
publisher = {Springer-Verlag},
address = {Berlin, Heidelberg}
}

@inproceedings{10.1007/978-3-030-79876-5_37,
author = {Moura, Leonardo de and Ullrich, Sebastian},
title = {The Lean 4 Theorem Prover and Programming Language},
year = {2021},
isbn = {978-3-030-79875-8},
publisher = {Springer-Verlag},
address = {Berlin, Heidelberg},
url = {https://doi.org/10.1007/978-3-030-79876-5_37},
doi = {10.1007/978-3-030-79876-5_37},
booktitle = {Automated Deduction – CADE 28: 28th International Conference on Automated Deduction, Virtual Event, July 12–15, 2021, Proceedings},
pages = {625–635},
numpages = {11}
}

@inproceedings{10.1145/3759425.3763384,
author = {Cao, Qinxiang and Xie, Lihan and Yan, Junchi},
title = {The LLM Era Demands Natural-Language-Aligned Theorem Provers for Mathematics},
year = {2025},
isbn = {9798400721489},
publisher = {Association for Computing Machinery},
address = {New York, NY, USA},
url = {https://doi.org/10.1145/3759425.3763384},
doi = {10.1145/3759425.3763384},
booktitle = {Proceedings of the 1st ACM SIGPLAN International Workshop on Language Models and Programming Languages},
pages = {46–50},
numpages = {5},
location = {Singapore, Singapore},
series = {LMPL '25}
}

@misc{liu2025theoremprovingformulationframework,
      title={Beyond Theorem Proving: Formulation, Framework and Benchmark for Formal Problem-Solving}, 
      author={Qi Liu and Xinhao Zheng and Renqiu Xia and Xingzhi Qi and Qinxiang Cao and Junchi Yan},
      year={2025},
      eprint={2505.04528},
      archivePrefix={arXiv},
      primaryClass={cs.AI},
      url={https://arxiv.org/abs/2505.04528}, 
}

@inproceedings{10.5555/3737916.3740575,
author = {Jiang, Albert Q. and Li, Wenda and Jamnik, Mateja},
title = {Multi-language diversity benefits autoformalization},
year = {2024},
isbn = {9798331314385},
publisher = {Curran Associates Inc.},
address = {Red Hook, NY, USA},
booktitle = {Proceedings of the 38th International Conference on Neural Information Processing Systems},
articleno = {2659},
numpages = {27},
location = {Vancouver, BC, Canada},
series = {NIPS '24}
}

@misc{wang2025kiminaproverpreviewlargeformal,
      title={Kimina-Prover Preview: Towards Large Formal Reasoning Models with Reinforcement Learning}, 
      author={Haiming Wang and Mert Unsal and Xiaohan Lin and Mantas Baksys and Junqi Liu and Marco Dos Santos and Flood Sung and Marina Vinyes and Zhenzhe Ying and Zekai Zhu and Jianqiao Lu and Hugues de Saxcé and Bolton Bailey and Chendong Song and Chenjun Xiao and Dehao Zhang and Ebony Zhang and Frederick Pu and Han Zhu and Jiawei Liu and Jonas Bayer and Julien Michel and Longhui Yu and Léo Dreyfus-Schmidt and Lewis Tunstall and Luigi Pagani and Moreira Machado and Pauline Bourigault and Ran Wang and Stanislas Polu and Thibaut Barroyer and Wen-Ding Li and Yazhe Niu and Yann Fleureau and Yangyang Hu and Zhouliang Yu and Zihan Wang and Zhilin Yang and Zhengying Liu and Jia Li},
      year={2025},
      eprint={2504.11354},
      archivePrefix={arXiv},
      primaryClass={cs.AI},
      url={https://arxiv.org/abs/2504.11354}, 
}

@misc{lu2026automatedformalizationconceptualretrievalaugmented,
      title={Automated Formalization via Conceptual Retrieval-Augmented LLMs}, 
      author={Wangyue Lu and Lun Du and Sirui Li and Ke Weng and Haozhe Sun and Hengyu Liu and Minghe Yu and Tiancheng Zhang and Ge Yu},
      year={2026},
      eprint={2508.06931},
      archivePrefix={arXiv},
      primaryClass={cs.AI},
      url={https://arxiv.org/abs/2508.06931}, 
}

@inproceedings{
liu2025rethinking,
title={Rethinking and Improving Autoformalization: Towards a Faithful Metric and a Dependency Retrieval-based Approach},
author={Qi Liu and Xinhao Zheng and Xudong Lu and Qinxiang Cao and Junchi Yan},
booktitle={The Thirteenth International Conference on Learning Representations},
year={2025},
url={https://openreview.net/forum?id=hUb2At2DsQ}
}

@misc{guo2025autoformalizertoolfeedback,
      title={Autoformalizer with Tool Feedback}, 
      author={Qi Guo and Jianing Wang and Jianfei Zhang and Deyang Kong and Xiangzhou Huang and Xiangyu Xi and Wei Wang and Jingang Wang and Xunliang Cai and Shikun Zhang and Wei Ye},
      year={2025},
      eprint={2510.06857},
      archivePrefix={arXiv},
      primaryClass={cs.AI},
      url={https://arxiv.org/abs/2510.06857}, 
}

@inproceedings{Leang_2025,
   title={Theorem Prover as a Judge for Synthetic Data Generation},
   url={http://dx.doi.org/10.18653/v1/2025.acl-long.1448},
   DOI={10.18653/v1/2025.acl-long.1448},
   booktitle={Proceedings of the 63rd Annual Meeting of the Association for Computational Linguistics (Volume 1: Long Papers)},
   publisher={Association for Computational Linguistics},
   author={Leang, Joshua Ong Jun and Hong, Giwon and Li, Wenda and Cohen, Shay B},
   year={2025},
   pages={29941–29977} }

@inproceedings{10.1007/978-3-642-03359-9_5,
author = {Naumowicz, Adam and Korni\l{}owicz, Artur},
title = {A Brief Overview of Mizar},
year = {2009},
isbn = {9783642033582},
publisher = {Springer-Verlag},
address = {Berlin, Heidelberg},
url = {https://doi.org/10.1007/978-3-642-03359-9_5},
doi = {10.1007/978-3-642-03359-9_5},
booktitle = {Proceedings of the 22nd International Conference on Theorem Proving in Higher Order Logics},
pages = {67–72},
numpages = {6},
location = {Munich, Germany},
series = {TPHOLs '09}
}

@article{10.1007/s10817-017-9440-6,
author = {Bancerek, Grzegorz and Byli\'{n}ski, Czes\l{}aw and Grabowski, Adam and Korni\l{}owicz, Artur and Matuszewski, Roman and Naumowicz, Adam and P\k{a}k, Karol},
title = {The Role of the Mizar Mathematical Library for Interactive Proof Development in Mizar},
year = {2018},
issue_date = {June      2018},
publisher = {Springer-Verlag},
address = {Berlin, Heidelberg},
volume = {61},
number = {1–4},
issn = {0168-7433},
url = {https://doi.org/10.1007/s10817-017-9440-6},
doi = {10.1007/s10817-017-9440-6},
journal = {J. Autom. Reason.},
month = jun,
pages = {9–32},
numpages = {24}
}

@inproceedings{The_mathlib_Community_2020, series={POPL ’20},
   title={The lean mathematical library},
   url={http://dx.doi.org/10.1145/3372885.3373824},
   DOI={10.1145/3372885.3373824},
   booktitle={Proceedings of the 9th ACM SIGPLAN International Conference on Certified Programs and Proofs},
   publisher={ACM},
   author={The mathlib Community},
   year={2020},
   month=Jan, pages={367–381},
   collection={POPL ’20} }

@inproceedings{
ospanov2026minifflean,
title={miniF2F-Lean Revisited: Reviewing Limitations and Charting a Path Forward},
author={Azim Ospanov and Farzan Farnia and Roozbeh Yousefzadeh},
booktitle={The Thirty-ninth Annual Conference on Neural Information Processing Systems},
year={2026},
url={https://openreview.net/forum?id=KtaHv0YUyh}
}

@misc{liu2025generalizedtreeeditdistance,
      title={Generalized Tree Edit Distance (GTED): A Faithful Evaluation Metric for Statement Autoformalization}, 
      author={Yuntian Liu and Tao Zhu and Xiaoyang Liu and Yu Chen and Zhaoxuan Liu and Qingfeng Guo and Jiashuo Zhang and Kangjie Bao and Tao Luo},
      year={2025},
      eprint={2507.07399},
      archivePrefix={arXiv},
      primaryClass={cs.LG},
      url={https://arxiv.org/abs/2507.07399}, 
}

@misc{zheng2022minif2fcrosssystembenchmarkformal,
      title={MiniF2F: a cross-system benchmark for formal Olympiad-level mathematics}, 
      author={Kunhao Zheng and Jesse Michael Han and Stanislas Polu},
      year={2022},
      eprint={2109.00110},
      archivePrefix={arXiv},
      primaryClass={cs.AI},
      url={https://arxiv.org/abs/2109.00110}, 
}

@misc{lin2025goedelproverfrontiermodelopensource,
      title={Goedel-Prover: A Frontier Model for Open-Source Automated Theorem Proving}, 
      author={Yong Lin and Shange Tang and Bohan Lyu and Jiayun Wu and Hongzhou Lin and Kaiyu Yang and Jia Li and Mengzhou Xia and Danqi Chen and Sanjeev Arora and Chi Jin},
      year={2025},
      eprint={2502.07640},
      archivePrefix={arXiv},
      primaryClass={cs.LG},
      url={https://arxiv.org/abs/2502.07640}, 
}

@misc{xin2024deepseekproveradvancingtheoremproving,
      title={DeepSeek-Prover: Advancing Theorem Proving in LLMs through Large-Scale Synthetic Data}, 
      author={Huajian Xin and Daya Guo and Zhihong Shao and Zhizhou Ren and Qihao Zhu and Bo Liu and Chong Ruan and Wenda Li and Xiaodan Liang},
      year={2024},
      eprint={2405.14333},
      archivePrefix={arXiv},
      primaryClass={cs.AI},
      url={https://arxiv.org/abs/2405.14333}, 
}

@article{Hubert2025,
  author    = {Thomas Hubert and Rishi Mehta and Laurent Sartran and Mikl{\'o}s Z. Horv{\'a}th and Goran {\v{Z}}u{\v{z}}i{\'c} and Eric Wieser and Aja Huang and Julian Schrittwieser and Yannick Schroecker and Hussain Masoom and Ottavia Bertolli and Tom Zahavy and Amol Mandhane and Jessica Yung and Iuliya Beloshapka and Borja Ibarz and Vivek Veeriah and Lei Yu and Oliver Nash and Paul Lezeau and Salvatore Mercuri and Calle S{\"o}nne and Bhavik Mehta and Alex Davies and Daniel Zheng and Fabian Pedregosa and Yin Li and Ingrid von Glehn and Mark Rowland and Samuel Albanie and Ameya Velingker and Simon Schmitt and Edward Lockhart and Edward Hughes and Henryk Michalewski and Nicolas Sonnerat and Demis Hassabis and Pushmeet Kohli and David Silver},
  title     = {Olympiad-level formal mathematical reasoning with reinforcement learning},
  journal   = {Nature},
  year      = {2025},
  month     = nov,
  day       = {12},
  doi       = {10.1038/s41586-025-09833-y},
  url       = {https://doi.org/10.1038/s41586-025-09833-y},
  issn      = {1476-4687}
}

@misc{wang2023legoproverneuraltheoremproving,
      title={LEGO-Prover: Neural Theorem Proving with Growing Libraries}, 
      author={Haiming Wang and Huajian Xin and Chuanyang Zheng and Lin Li and Zhengying Liu and Qingxing Cao and Yinya Huang and Jing Xiong and Han Shi and Enze Xie and Jian Yin and Zhenguo Li and Heng Liao and Xiaodan Liang},
      year={2023},
      eprint={2310.00656},
      archivePrefix={arXiv},
      primaryClass={cs.AI},
      url={https://arxiv.org/abs/2310.00656}, 
}

@misc{chen2025seedproverdeepbroadreasoning,
      title={Seed-Prover: Deep and Broad Reasoning for Automated Theorem Proving}, 
      author={Luoxin Chen and Jinming Gu and Liankai Huang and Wenhao Huang and Zhicheng Jiang and Allan Jie and Xiaoran Jin and Xing Jin and Chenggang Li and Kaijing Ma and Cheng Ren and Jiawei Shen and Wenlei Shi and Tong Sun and He Sun and Jiahui Wang and Siran Wang and Zhihong Wang and Chenrui Wei and Shufa Wei and Yonghui Wu and Yuchen Wu and Yihang Xia and Huajian Xin and Fan Yang and Huaiyuan Ying and Hongyi Yuan and Zheng Yuan and Tianyang Zhan and Chi Zhang and Yue Zhang and Ge Zhang and Tianyun Zhao and Jianqiu Zhao and Yichi Zhou and Thomas Hanwen Zhu},
      year={2025},
      eprint={2507.23726},
      archivePrefix={arXiv},
      primaryClass={cs.AI},
      url={https://arxiv.org/abs/2507.23726}, 
}

@misc{ren2025deepseekproverv2advancingformalmathematical,
      title={DeepSeek-Prover-V2: Advancing Formal Mathematical Reasoning via Reinforcement Learning for Subgoal Decomposition}, 
      author={Z. Z. Ren and Zhihong Shao and Junxiao Song and Huajian Xin and Haocheng Wang and Wanjia Zhao and Liyue Zhang and Zhe Fu and Qihao Zhu and Dejian Yang and Z. F. Wu and Zhibin Gou and Shirong Ma and Hongxuan Tang and Yuxuan Liu and Wenjun Gao and Daya Guo and Chong Ruan},
      year={2025},
      eprint={2504.21801},
      archivePrefix={arXiv},
      primaryClass={cs.CL},
      url={https://arxiv.org/abs/2504.21801}, 
}

@inproceedings{10.1007/978-3-030-79876-5_36,
author = {De Lon, Adrian and Koepke, Peter and Lorenzen, Anton and Marti, Adrian and Sch\"{u}tz, Marcel and Wenzel, Makarius},
title = {The Isabelle/Naproche Natural Language Proof Assistant},
year = {2021},
isbn = {978-3-030-79875-8},
publisher = {Springer-Verlag},
address = {Berlin, Heidelberg},
url = {https://doi.org/10.1007/978-3-030-79876-5_36},
doi = {10.1007/978-3-030-79876-5_36},
booktitle = {Automated Deduction – CADE 28: 28th International Conference on Automated Deduction, Virtual Event, July 12–15, 2021, Proceedings},
pages = {614–624},
numpages = {11}
}

@misc{jiang2023draftsketchproveguiding,
      title={Draft, Sketch, and Prove: Guiding Formal Theorem Provers with Informal Proofs}, 
      author={Albert Q. Jiang and Sean Welleck and Jin Peng Zhou and Wenda Li and Jiacheng Liu and Mateja Jamnik and Timothée Lacroix and Yuhuai Wu and Guillaume Lample},
      year={2023},
      eprint={2210.12283},
      archivePrefix={arXiv},
      primaryClass={cs.AI},
      url={https://arxiv.org/abs/2210.12283}, 
}

@inproceedings{wenzel1999isar,
  title     = {Isabelle/Isar: A Versatile Environment for Human-Readable Formal Proof Documents},
  author    = {Wenzel, Markus},
  booktitle = {Theorem Proving in Higher Order Logics: TPHOLs 1999},
  series    = {Lecture Notes in Computer Science},
  volume    = {1690},
  publisher = {Springer},
  year      = {1999}
}

@InProceedings{massot:LIPIcs.ITP.2024.27,
  author =	{Massot, Patrick},
  title =	{{Teaching Mathematics Using Lean and Controlled Natural Language}},
  booktitle =	{15th International Conference on Interactive Theorem Proving (ITP 2024)},
  pages =	{27:1--27:19},
  series =	{Leibniz International Proceedings in Informatics (LIPIcs)},
  ISBN =	{978-3-95977-337-9},
  ISSN =	{1868-8969},
  year =	{2024},
  volume =	{309},
  editor =	{Bertot, Yves and Kutsia, Temur and Norrish, Michael},
  publisher =	{Schloss Dagstuhl -- Leibniz-Zentrum f{\"u}r Informatik},
  address =	{Dagstuhl, Germany},
  URL =		{https://drops.dagstuhl.de/entities/document/10.4230/LIPIcs.ITP.2024.27},
  URN =		{urn:nbn:de:0030-drops-207550},
  doi =		{10.4230/LIPIcs.ITP.2024.27}
}

@book{demidovich1970problems,
  author    = {Demidovich, B. P.},
  title     = {Problems in Mathematical Analysis},
  publisher = {Mir Publishers},
  year      = {1970}
}

@misc{hu2021loralowrankadaptationlarge,
      title={LoRA: Low-Rank Adaptation of Large Language Models}, 
      author={Edward J. Hu and Yelong Shen and Phillip Wallis and Zeyuan Allen-Zhu and Yuanzhi Li and Shean Wang and Lu Wang and Weizhu Chen},
      year={2021},
      eprint={2106.09685},
      archivePrefix={arXiv},
      primaryClass={cs.CL},
      url={https://arxiv.org/abs/2106.09685}, 
}

@inproceedings{10.5555/1792233.1792261,
author = {Wiedijk, Freek},
title = {Mizar's soft type system},
year = {2007},
isbn = {3540745904},
publisher = {Springer-Verlag},
address = {Berlin, Heidelberg},
booktitle = {Proceedings of the 20th International Conference on Theorem Proving in Higher Order Logics},
pages = {383–399},
numpages = {17},
location = {Kaiserslautern, Germany},
series = {TPHOLs'07}
}

@misc{yang2025qwen3technicalreport,
      title={Qwen3 Technical Report}, 
      author={An Yang and Anfeng Li and Baosong Yang and Beichen Zhang and Binyuan Hui and Bo Zheng and Bowen Yu and Chang Gao and Chengen Huang and Chenxu Lv and Chujie Zheng and Dayiheng Liu and Fan Zhou and Fei Huang and Feng Hu and Hao Ge and Haoran Wei and Huan Lin and Jialong Tang and Jian Yang and Jianhong Tu and Jianwei Zhang and Jianxin Yang and Jiaxi Yang and Jing Zhou and Jingren Zhou and Junyang Lin and Kai Dang and Keqin Bao and Kexin Yang and Le Yu and Lianghao Deng and Mei Li and Mingfeng Xue and Mingze Li and Pei Zhang and Peng Wang and Qin Zhu and Rui Men and Ruize Gao and Shixuan Liu and Shuang Luo and Tianhao Li and Tianyi Tang and Wenbiao Yin and Xingzhang Ren and Xinyu Wang and Xinyu Zhang and Xuancheng Ren and Yang Fan and Yang Su and Yichang Zhang and Yinger Zhang and Yu Wan and Yuqiong Liu and Zekun Wang and Zeyu Cui and Zhenru Zhang and Zhipeng Zhou and Zihan Qiu},
      year={2025},
      eprint={2505.09388},
      archivePrefix={arXiv},
      primaryClass={cs.CL},
      url={https://arxiv.org/abs/2505.09388}, 
}

@misc{yang2024qwen2technicalreport,
      title={Qwen2 Technical Report}, 
      author={An Yang and Baosong Yang and Binyuan Hui and Bo Zheng and Bowen Yu and Chang Zhou and Chengpeng Li and Chengyuan Li and Dayiheng Liu and Fei Huang and Guanting Dong and Haoran Wei and Huan Lin and Jialong Tang and Jialin Wang and Jian Yang and Jianhong Tu and Jianwei Zhang and Jianxin Ma and Jianxin Yang and Jin Xu and Jingren Zhou and Jinze Bai and Jinzheng He and Junyang Lin and Kai Dang and Keming Lu and Keqin Chen and Kexin Yang and Mei Li and Mingfeng Xue and Na Ni and Pei Zhang and Peng Wang and Ru Peng and Rui Men and Ruize Gao and Runji Lin and Shijie Wang and Shuai Bai and Sinan Tan and Tianhang Zhu and Tianhao Li and Tianyu Liu and Wenbin Ge and Xiaodong Deng and Xiaohuan Zhou and Xingzhang Ren and Xinyu Zhang and Xipin Wei and Xuancheng Ren and Xuejing Liu and Yang Fan and Yang Yao and Yichang Zhang and Yu Wan and Yunfei Chu and Yuqiong Liu and Zeyu Cui and Zhenru Zhang and Zhifang Guo and Zhihao Fan},
      year={2024},
      eprint={2407.10671},
      archivePrefix={arXiv},
      primaryClass={cs.CL},
      url={https://arxiv.org/abs/2407.10671}, 
}

\clearpage
\appendix

\section{ An example proof written in the Relaxed NFL } \label{app:A}

This appendix presents the Relaxed NFL formalization of a functional-equation problem from the 1999 International Mathematical Olympiad.
The solution is adapted from the one available on the Art of Problem Solving website:
\url{https://artofproblemsolving.com/wiki/index.php/1999_IMO_Problems/Problem_6}.

\begin{tcolorbox}
\begin{lstlisting}[keepspaces=true]
Solve: Find all f such that
  IsFunc(f),
  f : Real -> Real,
  forall (x, y in Real),
    f(x - f(y)) = f(f(y)) + x * f(y) + f(x) - 1

Solution:
  Set f(0) = c
  
  [@method Substitute x = y = 0 into forall (x, y in Real),
    f(x - f(y)) = f(f(y)) + x * f(y) + f(x) - 1 @]
  We have f(-c) = f(c) + c - 1

  Assume c = 0 holds {
    We have f(0) = f(0) - 1
    Contradiction
  }

  We have c != 0

  [@method Substitute x = f(y) into forall (x, y in Real),
    f(x - f(y)) = f(f(y)) + x * f(y) + f(x) - 1 @]
  We have c = f(x) + x ^ 2 + f(x) - 1

  Solve the system of equations c = f(x) + x ^ 2 + f(x) - 1 {
    We have f(x) = frac(c + 1, 2) - frac(x ^ 2, 2)
  }

  [@method By x ^ 2 = (-x) ^ 2 @]
  We have f(x) = f(-x)
  We have f(c) = f(-c)
  We have f(c) = f(c) + c - 1
  We have c = 1
  We have f(x) = 1 - frac(x ^ 2, 2)
\end{lstlisting}
\end{tcolorbox}

\clearpage

\section{ Grammar of the Relaxed Natural Formal Language } \label{app:B}

\begin{grammar}

<program> ::= <assumptions> "Prove:" <term> "Proof:" <proof>
\alt <assumptions> "Solve:" <solving\_step> "Solution:" <derivation>

<assumptions> ::= "Assume:" <term> ("," <term>)*

<proof> ::= <derivation> "QED"

<derivation> ::= <annotated\_step>*

<annotated\_step> ::= ("[@method" <method> "@]")? <step>

<step> ::= <forward\_step>
\alt <backward\_step>
\alt <subgoal\_step> "{" <proof> "}"
\alt <action\_step>
\alt <local\_action\_step> "{" <arguments> "}"
\alt <solving\_step> "{" <arguments> "}"

<forward\_step> ::= "We have" <term>
\alt "Contradiction"

<backward\_step> ::= "It suffices to prove" <term>

<subgoal\_step> ::= "We prove that" <term>

<action\_step> ::= "Set" <term> "=" <term>
\alt "Take" <term> "as a witness"
\alt "After checking," <term> ("," <term>) "satisfies the constraints"
\alt "After checking," <term> ("," <term>) "does not satisfy the constraints"

<local\_action\_step> ::= "Assume" <term> ("," <term>) "holds"
\alt "For arbitrary" <term> ("," <term>) "satisfying" <term> ("," <term>)
\alt "In order for" <term> "to hold"

<solving\_step> ::= "Find" <term> ("," <term>) "such that" <term> ("," <term>)
\alt "Find all" <term> ("," <term>) "such that" <term> ("," <term>)
\alt "Find the range of" <term> ("," <term>) "such that" <term> ("," <term>)
\alt "Solve the system of equations" <term> ("," <term>)
\alt "Solve the system of inequalities" <term> ("," <term>)
\alt "Compute" <term> ("," <term>)

<method> ::= "By" <term>
\alt "By" <string>
\alt "Similarly"
\alt "Substitute" <term> ("," <term>) "into" <term>
\alt "Multiply both sides of" <term> "by" <term>
\alt "Square both sides of" <term>
\alt "Take limits on both sides" <term> "as" <term>
\alt ...

\end{grammar}

\end{document}